\documentclass[preprint,aps,tightenlines,floatfix,showpacs]{revtex4}
\usepackage{graphics}
\usepackage{bm}
\begin{document}
\preprint{LA-UR-02-834}
\title{Total reaction cross sections for neutron-nucleus scattering}
\author{K. Amos}
\email{amos@physics.unimelb.edu.au}
\affiliation{School of Physics, University of Melbourne, 
Victoria 3010, Australia}
\author{S. Karataglidis}
\email{stevenk@lanl.gov}
\affiliation{Theoretical Division, Los Alamos National
Laboratory, Los Alamos, New Mexico, 87545}
\date{\today}
\begin{abstract}
Neutron total reaction cross sections at 45, 50, 55, 60, 65, and 75
MeV from nuclei $^{12}$C, $^{28}$Si, $^{56}$Fe, $^{90}$Zr, and
$^{208}$Pb have been measured and are compared with (microscopic)
optical model predictions. The optical potentials were obtained in
coordinate space by full folding effective nucleon-nucleon
interactions with realistic nuclear ground state density
matrices. Good to excellent agreement is found.
\end{abstract}
\pacs{25.40.-h,24.10.Ht,21.60.Cs,24.10.Eq}
\maketitle

The usual vehicle for specifying $NA$ total reaction cross sections,
required as inputs in diverse applications ranging from transmutation
of waste to nuclear astrophysics, has been the $NA$ optical potential;
a potential most commonly taken as a local parametrized function,
usually of Woods-Saxon type.  However, it has long been known that the
optical potential must be nonlocal and markedly so, although it has
been assumed also that the energy dependence of the customary
(phenomenological) model accounts for that.  Of more concern is that
the phenomenological approach is not truly predictive.  The parameter
values chosen, while they may be set from a global survey of data
analyses, are subject to uncertainties and ambiguities, stemming in
part from the lack of any density dependences of the specific
scattering systems. While differential cross sections may be described
by such local potentials, different parameter sets, describing the
same elastic scattering data, may lead to different predictions for
the total reaction cross section.

In this letter we compare predictions with new experimental data on
total reaction cross sections for neutron scattering at a number of
intermediate energies and from five nuclei ranging in mass from
$^{12}$C to $^{208}$Pb \cite{Ib02}; the results were compared with the
LA-150 data~\cite{Ch99}. Unlike those analyses using phenomenological
local form interactions, the predictions we report were found using,
without approximation, complex, nonlocal, coordinate-space optical
potentials formed by full folding realistic, effective,
nucleon-nucleon ($NN$) interactions with density matrices (hereafter
simply termed densities) specified from credible models of the
structure of the targets.  All details of this approach have been
given in a recent review~\cite{Am00}.  With this coordinate space
approach, all analyses have been made using the DWBA98 programs
\cite{Ra99} which allow the projectile-target nucleon interactions not
only to be complex but also energy and density dependent, the latter
specific to the target in question.  The properties of (nonlocal)
optical potentials arise from mapping effective interactions to $NN$
$g$ matrices that are solutions of the Bruckner-Bethe-Goldstone (BBG)
equations for nuclear matter.  We use the Bonn-B $NN$ potentials as
input to solving those BBG equations. As the BBG equations include
allowance for Pauli blocking and of medium influence in the
propagators, the ensuing effective $NN$ interactions that will be
folded with the nuclear structure are nuclear density dependent.  Such
density dependence has been shown to be critical in making good
predictions of angular and integral observables, including all spin
observables, for intermediate energy nucleon scattering from all
nuclei~\cite{Am00,De01}.

Formally, the nonlocal optical potentials can be written
\begin{eqnarray}
U(\bm{r}_1,\bm{r}_2; E) & = & \sum_n \zeta_n \left\{ \delta(\bm{r}_1
- \bm{r}_2) \int \varphi^*_n(\bm{s})\, v_D(\bm{r}_{1s})\,
\varphi_n(\bm{s}) \, d\bm{s} + \varphi^*_n(\bm{r}_1)\,
v_{Ex}(\bm{r}_{12})\, \varphi_n(\bm{r}_2) \right\} \nonumber\\
& \Rightarrow & U_D(\bm{r}_1; E) \delta(\bm{r}_1 - \bm{r}_2) +
U_{Ex}(\bm{r}_1,\bm{r}_2; E)\, ,
\label{NonSE}
\end{eqnarray}
where $v_D$ , $v_{Ex}$ are combinations of the components of the
effective $NN$ interactions, $\zeta_n$ are ground state one body
density matrices (which often reduce to bound state shell
occupancies), and $\varphi_n(\bm{r})$ are single nucleon bound
states.  All details and the prescription of solution of the
associated nonlocal Schr\"odinger equations are given in the recent
review~\cite{Am00}.

The specification of the nuclear ground state density is taken from a
given nucleon-based model of structure. For $^{12}$C we have used a
complete $(0+2)\hbar\omega$ shell model \cite{De01} using the WBT
interaction of Warburton and Brown \cite{Wa92}. For $^{28}$Si, we have
used a complete $0\hbar\omega$ shell model wave function as formed by
Brown and Warburton \cite{Br88}.  With $^{56}$Fe a packed shell model
was chosen with allowance of $2p-2h$ excitations into the $f-p$ shell.
Potentials used in the structure calculation were those of Richter
{\em et al.} \cite{Ri91}.  The NIS model of Ji and Wildenthal
\cite{Ji89} was used to describe $^{90}$Zr.  With that the protons are
in the packed shells while neutrons were allowed to spread in the
$f_{\frac{5}{2}}, p_{\frac{3}{2}}, p_{\frac{1}{2}}$, and
$g_{\frac{9}{2}}$ orbits.  With $^{208}$Pb we have used two recent
Skyrme-Hartree-Fock wave functions~\cite{Ba00,Ka02} for which the
neutron skin is 0.17~fm.

With $E \propto k^2$, the total reaction cross sections are 
\begin{equation}
\sigma_R(E)  =  \frac{\pi}{k^2} \sum^{\infty}_{l=0} \left\{ (l+1)
\left[ 1 - \left( \eta^+_l \right)^2 \right] + l \left[ 1 - \left(
\eta^-_l \right)^2 \right] \right\} \;,
\end{equation}
where $\eta^\pm_l $ are the moduli of $S$ matrices that are determined
from solutions of the radial Sch\"odinger equations with the specified
optical potentials. They are
\begin{equation}
\eta^{\pm}_l \equiv \eta^{\pm}_l(k) = \left| S^{\pm}_l(k) \right| =
e^{-2\Im\left[ \delta^{\pm}_l(k) \right] } \;,
\end{equation}
where  $\delta^{\pm}_l(k)$ are the scattering phase shifts.

\begin{figure}
\includegraphics{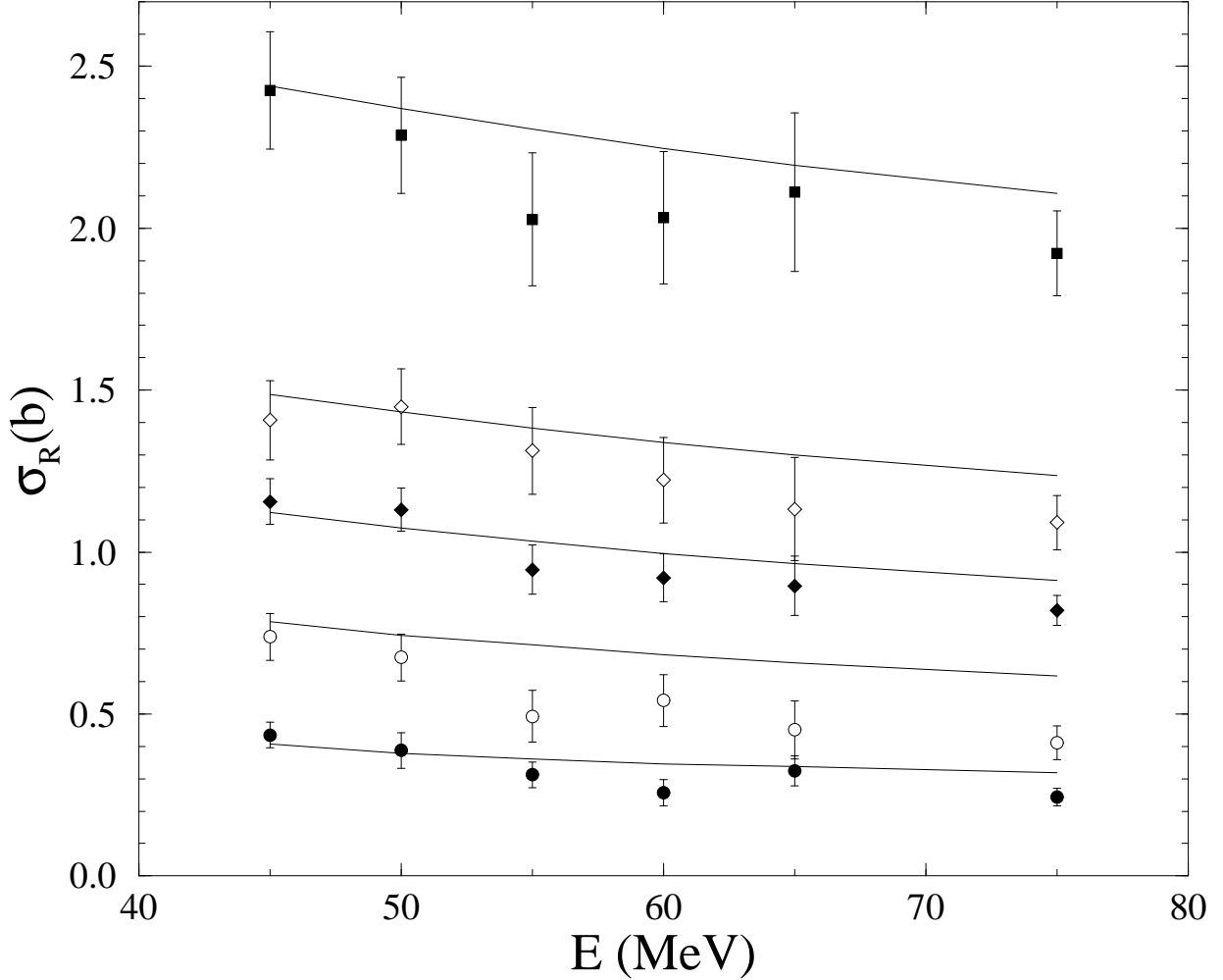}
\caption{\label{neut-A-sigr} Experimental values of $\sigma_R$ for
neutron scattering from $^{12}$C (closed circles), $^{28}$Si (open
circles), $^{56}$Fe (closed diamonds), $^{90}$Zr (open diamonds), and
$^{208}$Pb (squares).  The solid curves are lines connecting
predictions of our optical potential calculations.}
\label{figure}
\end{figure}

The results are given in Table~\ref{table} and also in Fig.~\ref{figure}.
The units are in barn.
\begin{table}
\begin{ruledtabular}
\caption{Neutron total reaction cross sections in barn.}
\label{table}
\begin{tabular}{ccccccl}
Energy (MeV) & $^{12}$C & $^{28}$Si & $^{56}$Fe & $^{90}$Zr & $^{208}$Pb & type \\
\hline
45 & 0.435 (39)   & 0.738 (72)   & 1.156 (71)  & 1.407 (122)  & 2.425 (181)  & expt.\\
   & 0.408        & 0.786        & 1.122       & 1.487        & 2.443        & SHF \\
   &            &            &            &             & 2439        & SKX \\
50 & 0.388 (55)   & 0.675 (72)   & 1.131 (67)  & 1.449 (117)  & 2.287 (179)  & expt.\\
   & 0.378        & 0.742        & 1.075       & 1.432        & 2.376        & SHF \\
   &            &            &            &             & 2370        & SKX\\
55 & 0.312 (40)   & 0.493 (80)   & 0.946 (76)   & 1.313 (134)  & 2.027 (205)  & expt.\\
   & 0.362        & 0.713        & 1.033       & 1.382        & 2.315        & SHF \\
   &            &            &            &             & 2306        & SKX \\
60 & 0.257 (40)   & 0.542 (80)   & 0.921 (75)   & 1.222 (132)  & 2.032 (204)  & expt.\\
   & 0.346        & 0.684        & 0.996        & 1.338        & 2.277        & SHF \\
   &            &            &            &             & 2247        & SKX \\
65 & 0.325 (46) & 0.451 (89) & 0.896 (91) & 1.133 (159) & 2.112 (245) & expt.\\
   & 0.338      & 0.658      & 0.964      & 1.299       & 2.226       & SHF \\
   &            &            &            &             & 2.195       & SKX\\
75 & 0.244 (27) & 0.412 (52) & 0.820 (47) & 1.091  (84) & 1.922 (131) & expt.\\
   & 0.319      & 0.617      & 0.913      & 1.236       & 2.075       & SHF\\
   &            &            &            &             & 2.108       & SKX\\
\end{tabular}
\end{ruledtabular}
\end{table}
From the table it is clear that our predictions compare well with most
of the data with the exception of the 55~MeV values and of those from
$^{28}$Si in particular.  Also there is a tendency for our
predictions to be higher than the data for all five nuclei at 75 MeV.
Those effects are revealed also in the figure.
However, inspection of the plotted results stresses that the
statistical error bars encompass most of our predicted values. While
in general the energy trends are well reproduced, we note that there
appears to be a discrepancy in the case of $^{28}$Si, where our
results tend to overpredict the data.

We have used a microscopic, nonlocal, optical model to predict total
reaction cross sections for the scattering of intermediate energy
neutrons from various nuclei. The agreement between the set of results
found for neutron reaction cross sections and the new data is now
comparable with that obtained for proton scattering \cite{De01}. We
note that in that earlier work the available neutron reaction cross
sections were consistently overpredicted by $\sim 10$\%. The data used
in that paper were far older and relied on the subtraction of the
large elastic scattering cross sections from the total cross
sections. The level of agreement found with the present data are much
better and give confidence in the predicted results. Thus the nonlocal
optical potentials generated by full folding $NN$ $g$ matrices with
microscopic (nucleon) structures for targets can be used to predict
both neutron and proton reaction cross sections.

\begin{acknowledgments}
This work was supported by a grant from the Australian Research
Council and also by DOE Contract no. W-7405-ENG-36.
\end{acknowledgments}

\bibliography{neut-A.bib}

\end{document}